\documentclass[11pt, twoside]{article}   	
\usepackage[margin=1in]{geometry}
\pdfoutput=1
\usepackage{amsmath,graphicx,siunitx} 
\usepackage[hypcap]{caption}
\usepackage{setspace}
\usepackage{newfloat}
\usepackage{authblk}
\usepackage[lowtilde]{url}
\newcommand{\beginsupplement}{%
        \setcounter{table}{0}
        \renewcommand{\thetable}{S\arabic{table}}%
        \setcounter{figure}{0}
        \renewcommand{\thefigure}{S\arabic{figure}}%
     }

\title{Bipartite Community Structure of eQTLs}
\author[1,2]{John Platig\thanks{jplatig@jimmy.harvard.edu}\textsuperscript{,}}
\author[3,4,5,6]{Peter Castaldi}
\author[3,5,6]{Dawn DeMeo}
\author[1,2,3]{John Quackenbush}
\affil[1]{Department of Biostatistics and Computational Biology, Dana-Farber Cancer Institute, Boston, MA}
\affil[2]{Department of Biostatistics, Harvard Chan School of Public Health, Boston, MA}
\affil[3]{Channing Division of Network Medicine, Brigham and Women's Hospital, Boston, MA}
\affil[4]{Division of General Medicine, Brigham and Women's Hospital, Boston, MA}
\affil[5]{Division of Pulmonary and Critical Care Medicine, Brigham and Women's Hospital, Boston, MA}
\affil[6]{Harvard Medical School, Boston, MA}
\begin{document}
\maketitle
\section{Abstract}
{\bf Keywords:} Genomics, GWAS, eQTL, COPD, Bipartite Networks, Community Structure \\[5mm]
Genome Wide Association Studies (GWAS) and eQTL analyses have produced a large and growing number of genetic associations linked to a wide range of human phenotypes. As of 2013, there were more than 11,000 SNPs associated with a trait as reported in the NHGRI GWAS Catalog. However, interpreting the functional roles played by these SNPs remains a challenge. Here we describe an approach that uses the inherent bipartite structure of eQTL networks to place SNPs into a functional context.

Using genotyping and gene expression data from 163 lung tissue samples in a study of Chronic Obstructive Pulmonary Disease (COPD) we calculated eQTL associations between SNPs and genes and cast significant associations (FDR $< 0.1$) as links in a bipartite network. To our surprise, we discovered that the highly-connected ``hub''  SNPs within the network were devoid of disease-associations. However, within the network we identified 35 highly modular communities, which comprise groups of SNPs associated with groups of genes; 13 of these communities were significantly enriched for distinct biological functions (P $ < \num{5e-4}$) including COPD-related functions. Further, we found that GWAS-significant SNPs were enriched at the cores of these communities, including previously identified GWAS associations for COPD, asthma, and pulmonary function, among others. These results speak to our intuition: rather than single SNPs influencing single genes, we see groups of SNPs associated with the expression of families of functionally related genes and that disease SNPs are associated with the perturbation of those functions. These methods are not limited in their application to COPD and can be used in the analysis of a wide variety of disease processes and other phenotypic traits.

\section{Introduction}
Genome Wide Association Studies (GWAS) have created new opportunities to understand the genetic factors that influence complex traits. Excepting  highly-penetrant Mendelian disorders, the majority of genetic associations seem to be driven by many factors, each of which has a relatively small effect. In a recent study \cite{wood2014}, 697 SNPs were associated with height in humans at genome-wide significance, yet these SNPs were able to explain only $\sim$20\% of height variability; $\sim$9,500 SNPs were needed to raise that to $\sim$29\%. In addition, $\sim 95\%$ of GWAS variants map to non-coding regions \cite{ardlie2015genotype}, complicating biological interpretation of their functional impact. 
\par
To bridge the functional gap between genetic variant and complex trait, expression Quantitative Trait Locus (eQTL) analysis associates SNP genotype with gene expression levels. Most eQTL analyses have focused on \emph{cis-}SNPs--those near the Transcriptional Start Site (TSS) of the gene in the association test. Recent computational developments \cite{shabalin2012} and work demonstrating the impact and replicability of \emph{trans}-eQTLs \cite{westra2013, fehrmann2011trans} have increased interest in identifying and understanding the role played by \emph{trans-}acting SNPs. 
\par
\begin{figure}[h!]
\centering
\includegraphics[trim = 1cm 1cm 1cm 3cm, clip, width=0.8\textwidth]{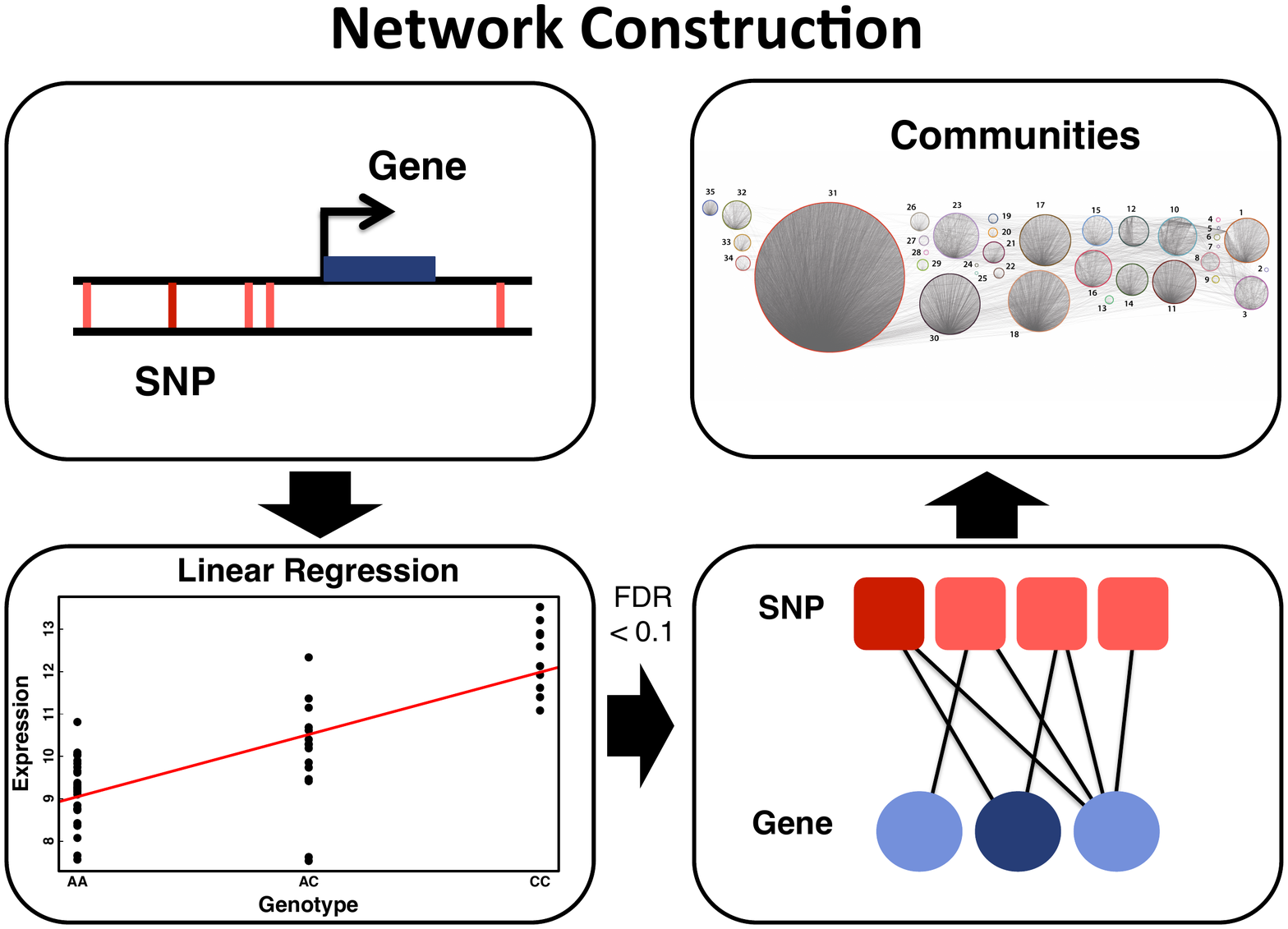}
\caption{Summary of the Network Method: All possible SNP-Gene pairs measured in the LGRC data set are considered in the eQTL analysis. Those SNP-gene pairs that are significantly associated (FDR $<0.1$) are included in the bipartite network. Communities were detected in the network using a bipartite modularity maximization approach, producing 35 highly modular communities, 13 of which are enriched for various GO terms. GWAS-associated SNPs are much more likely to lie at the cores of these communities.\label{fig:schematic}}
\end{figure}

However, new methods are needed to elucidate the potential functional impact of the thousands of GWAS and eQTL SNPs that can be detected in a single study. Here we present a complex networks method (Fig.\ \ref{fig:schematic}) that incorporates both \emph{cis-} and \emph{trans-} associations to identify groups of SNPs that are linked to groups of genes and systematically interrogate their biological functions. We then validate this approach using genotyping and gene expression data from 163 lung tissue samples in a study of Chronic Obstructive Pulmonary Disease (COPD) by the Lung Genomics Research Consortium (LGRC).

\section*{eQTL Networks}
\label{sec:eQTL}
We used the \texttt{MatrixeQTL} package in \texttt{R} to calculate all \emph{cis}- and \emph{trans}-eQTLs, considering only autosomal SNPs, using age, sex, and pack-years as covariates (see Supplementary Methods). The \emph{cis-} and \emph{trans-} associations were run separately, with an FDR threshold of 10\%. This analysis identified 32,053 \emph{cis}-eQTLs and 39,107 \emph{trans}-eQTLs. Quantile-quantile plots for both \emph{cis-} and \emph{trans-} are shown in Supplementary Figure \ref{fig:qqplot}. In total, 71,160 statistically significant associations were detected between 54,475 SNPs and 7,091 genes.
\par
\begin{figure}[h!]
\centering
\includegraphics[width=0.75\textwidth]{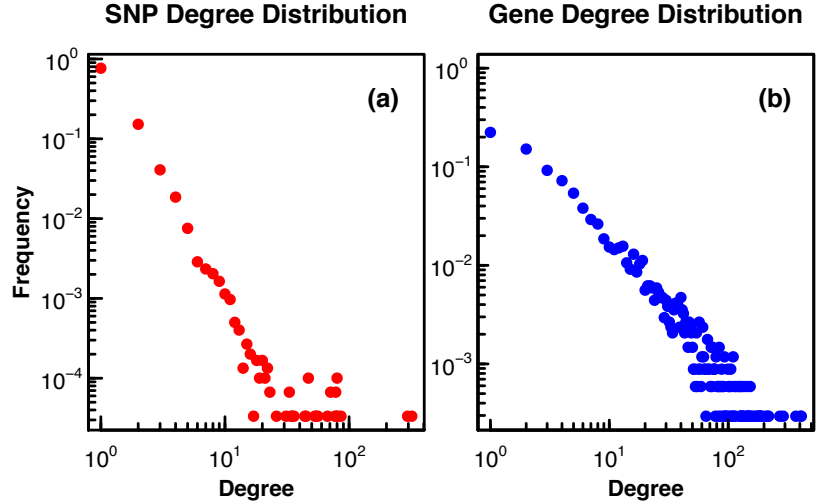}
\caption{Plot of the degree distribution, showing the frequency of node degree plotted on a log-log scale, for SNPs (left) and genes (right) in the giant connected component of the bipartite eQTL network.\label{fig:degree_dist}}
\end{figure}

We represented these associations as a bipartite network consisting of two classes of nodes---SNPs and genes---with edges from SNPs to the genes with which they are significantly associated based on the eQTL FDR cut-off. The network had a Giant Connected Component (GCC) with 44,872 links, 29,907 SNPs, and 3,390 genes. As a network diagnostic, we plotted the distributions of edges per SNP--called the SNP degree--and edges per gene (Fig.\ \ref{fig:degree_dist}) for the SNPs and genes in the GCC. The degree distributions for both the SNPs and genes are broad-tailed, implying potential power-law behavior. To test this, we fit each degree distribution to a power law, and determined the goodness of fit using the method described in \cite{clausetpower} (see Supplementary Methods). The probability the SNP degree follows a power-law distribution is very low, $P_{pl} \approx 0$, and the gene degree distribution (Fig.\ \ref{fig:degree_dist}b) is also unlikely to be power-law distributed ($P_{pl} < \num{0.1}$) even though there are multiple network hubs, shown in the tail of the distribution in Figure \ref{fig:degree_dist}b. 
\par
\begin{figure}[h!]
\centering
\includegraphics[width=0.5\textwidth]{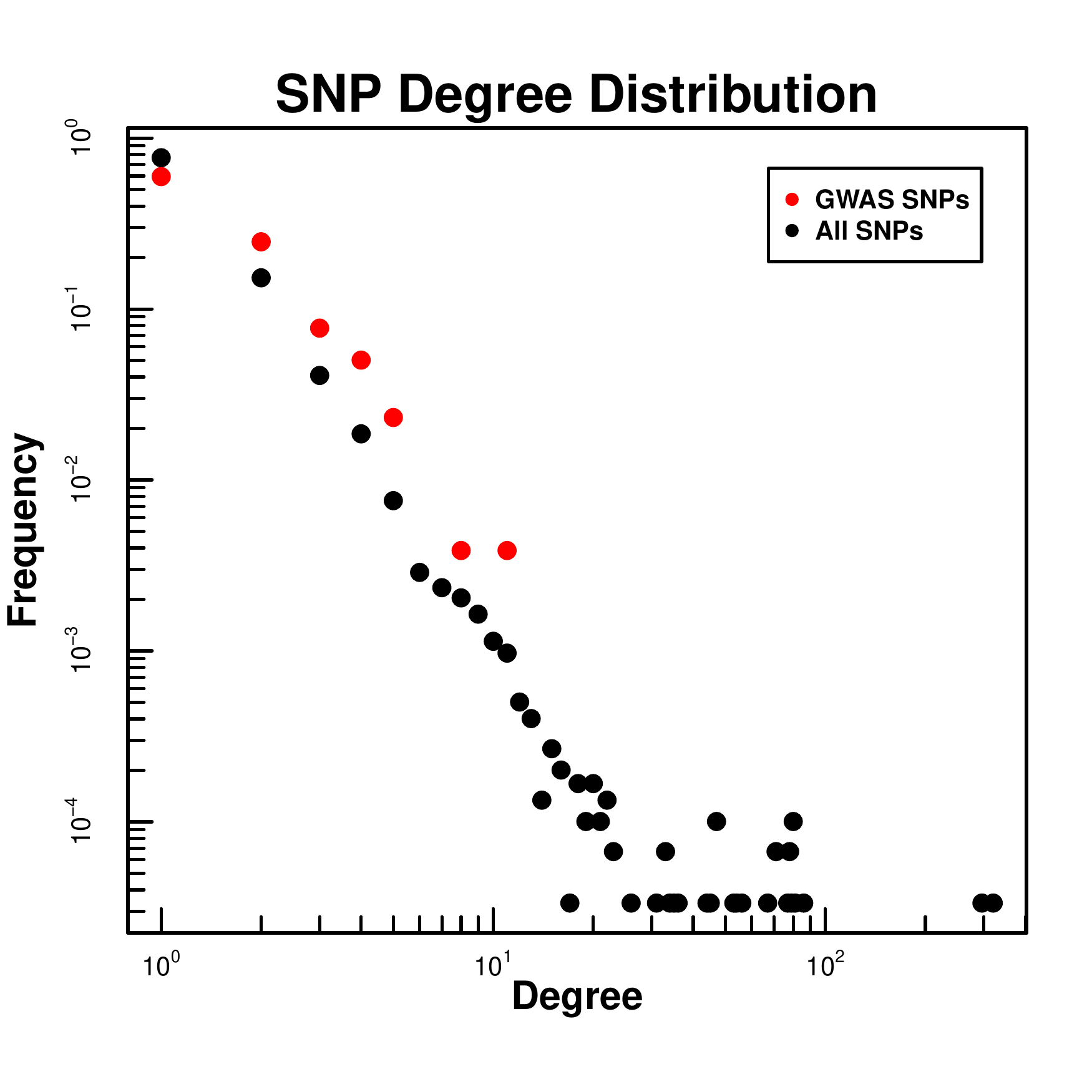}
\caption{Degree distributions for NHGRI-GWAS (red) and non-GWAS (black) SNPs. NHGRI-GWAS SNPs tend not to be network ``hubs,'' which are located in the far-right tail of the distribution. The highest degree NHGRI-GWAS SNP was connected to 11 genes. \label{fig:gwas_degree}}
\end{figure}

It is often cited in complex networks literature that the hubs, those nodes in the network that are most highly connected, represent critical elements whose removal can disrupt the entire network \cite{albert2000error, jeong2000large}. As a result, one widely-held belief about biological networks is that disease-related elements should be over-represented among the network hubs \cite{barabasihub}. To test the hypothesis that disease-associated SNPs are concentrated in the hubs, we projected GWAS-identified SNPs associated with a wide range of diseases and phenotypes onto the SNP degree distribution (Fig.\ \ref{fig:gwas_degree}). We used the \emph{gwascat} package \cite{gwascat} in \texttt{R} to download GWAS SNPs annotated in the NHGRI GWAS catalog; $259$ of those SNPs mapped to the GCC of the eQTL network. To our surprise, the network hubs--the right tail of Figure \ref{fig:gwas_degree}--were devoid of disease-associated SNPs which were instead scattered through the upper left half of the degree distribution. While the SNPs associated with a single gene are easier to interpret, the concentration of disease-associated SNPs in the middle of the distribution prompted us to look at other features of the network and its structure.

\subsection*{Community Structure Analysis}
\label{sec:community}

Many real-world networks have a complex structure consisting of ``communities'' of nodes \cite{girvanmodularity}. These communities are often defined as a group of network nodes that are more likely to be connected to other nodes within their community than they are to those outside of the community. A widely used measure of community structure is the modularity, which can be interpreted as an enrichment for links within communities minus an expected enrichment given the network degree distribution \cite{newmanmodularity}.
\par 
To partition the nodes from the eQTL network into communities---each of which contains both SNPs and genes---we maximized the bipartite modularity \cite{brim}. As recursive cluster identification and optimization can be computationally slow, we calculated an initial community structure assignment on the weighted, gene-space projection, using a fast uni-partite modularity maximization algorithm \cite{blondel2008fast} available in the \texttt{R} \emph{igraph} package \cite{igraph}, then iteratively converged ($\Delta Q < 10^{-4}$) on a community structure corresponding to a maximum bipartite modularity. 
\par	
The bipartite modularity is defined in Eq. (\ref{eq:bimod}), where $m$ is the number of links in the network, $\widetilde{A}_{ij}$ is the upper right block of the network adjacency matrix (a binary matrix where a 1 represents a connection between a SNP and a gene and 0 otherwise), $k_{i}$ is the degree of SNP $i$, $d_{j}$ is the degree of gene $j$, and $C_{i}$, $C_{j}$ the community indices of SNP $i$ and gene $j$, respectively. 

	\begin{equation} \label{eq:bimod}
	Q = \frac{1}{m}\sum_{i,j}\left(\widetilde{A}_{ij}-\frac{k_{i}d_{j}}{m}\right)\delta(C_{i},C_{j})
	\end{equation} 

\begin{figure}[h!]
\centering
\includegraphics[trim = 0cm 6cm 0cm 5cm, clip,width=0.8\textwidth]{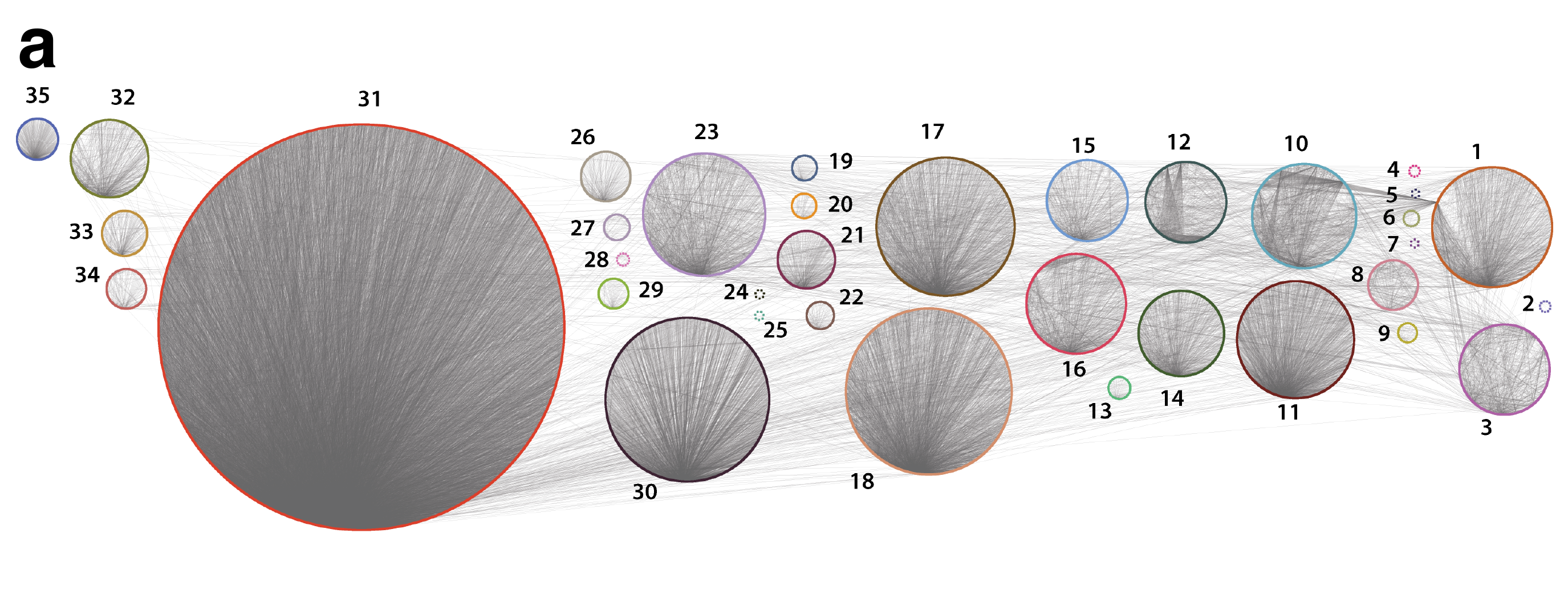}
\includegraphics[width=1\textwidth]{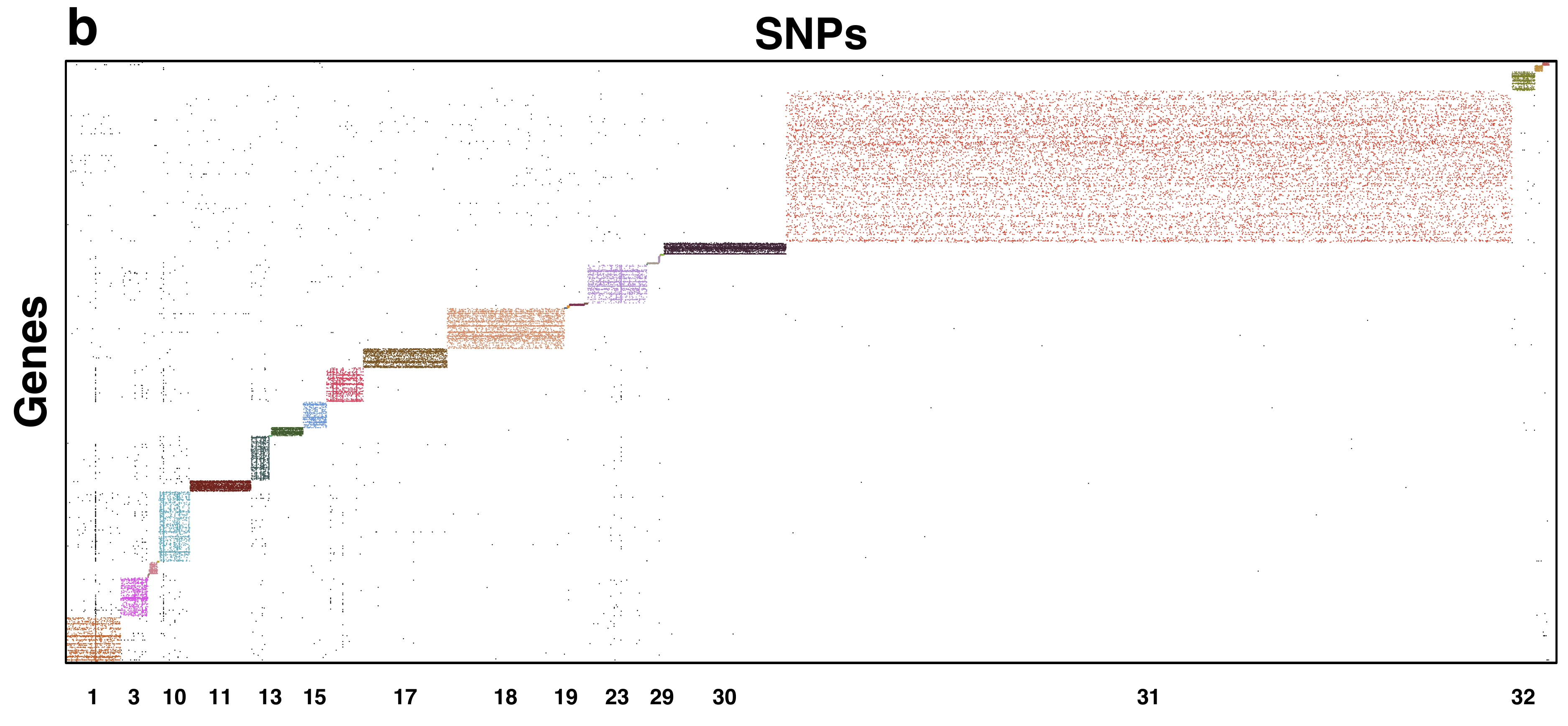} 
\caption{ {\bf (a)} Plot of the communities within the bipartite eQTL network. The nodes (genes and SNPs) in each community form a ring, with the link density within each ring visibly darker than links between communities. {\bf (b)} Links within communities (colored points) are shown along the diagonal, with links that go between communities in black. Community IDs are plotted along the $x$-axis.\label{fig:comms} }
\end{figure}

This analysis identified 35 highly modular communities in the LGRC data ($Q=0.766$; Fig.\ \ref{fig:comms}). The density of these communities can be seen in Figure \ref{fig:comms}. In Figure \ref{fig:comms}b, there is visible enrichment for links within each community (colored links) compared to links between different communities (black links). These communities represent groups of SNPs and genes that are highly connected to each other, suggesting that groups of genes may be jointly moderated by groups of SNPs that together represent specific biological processes. 
\par 
To investigate this hypothesis, we tested each community for GO term enrichment using Fisher's Exact Test (available in the \texttt{R} package \emph{GOstats} \cite{gostats}) and found 13 of the 35 communities contained genes enriched for specific Gene Ontology terms ($P< \num{5e-4}$; overlap $> 4$), encompassing a broad collection of cellular functions that are not generally associated with COPD. Indeed, this is what one might expect as the genetic background of an individual should have an effect not only on disease-specific processes, but more globally on the physiology of his or her individual cells. A number of communities do, however, show enrichment for biological processes that are known to be involved in COPD, including genes previously associated with the disease.

For example, Community 18 (see Fig.\ \ref{fig:comms}) was enriched for chromatin and nucleosome assembly/organization and includes members of the HIST1H gene superfamily. Community 30 (see Fig.\ \ref{fig:comms}) included GO term enrichment for function related to the HLA gene family, including T cell function and immune response; autoimmunity has been suggested as a potential contributor to COPD pathogenesis \cite{autocopd}. This community also contains PSORS1C1, which has been previously implicated in COPD \cite{qiu2011genetics}.
\par
Another of the genes in Community 30, AGER, has been implicated in COPD \cite{ager} and encodes sRAGE, a biomarker for emphysema. Its expression is negatively associated via eQTL analysis ($\beta = -0.3$) with rs6924102. This SNP has been observed to be an eQTL in a large blood eQTL dataset for a number of neighboring genes \cite{westra2013}, but it has not previously been described as an eQTL for AGER. This SNP lies in a region containing a DNase peak in cell lines analyzed by ENCODE \cite{encode2012} (indicating it sits in a region of open chromatin) and there is evidence of POLR2A binding from ChIP-Seq data in the GM12878 cell line as reported by ENCODE (http://regulomedb.org/snp/chr6/32811382). This suggests that rs6924102 may inhibit the expression of AGER through disruption of RNA Polymerase II binding and subsequent mRNA synthesis. This SNP is located $\sim$700KB from the well-studied non-synonymous AGER SNP, rs2070600.
\par
Examining Figure \ref{fig:comms}a, it is evident that within each community there are local hubs that are highly connected to the genes within that community. To identify these local hubs, we defined a ``core score'' that estimates importance of a SNP in the structure of its community. For SNP $i$ in community $h$, its core score, $Q_{ih}$, Eq.\ (\ref{eq:coreq}), is the fraction of the modularity of community $h$, $Q_{h}$, Eq.\ (\ref{eq:Qcom}), contributed by SNP $i$. This allows for comparison of SNPs from different communities, as each community does not have the same modularity, $Q_{h}$.
	\begin{align}
	Q_{ih} &= \frac{\frac{1}{m} \sum_{j}\left(\widetilde{A}_{ij}-\frac{k_{i}d_{j}}{m}\right)\delta(C_{i},h)\delta(C_{j},h)}{Q_{h}}	 \label{eq:coreq}	\\
	Q_{h} &= \frac{1}{m}\sum_{i,j}\left(\widetilde{A}_{ij}-\frac{k_{i}d_{j}}{m}\right)\delta(C_{i},h)\delta(C_{j},h)  \label{eq:Qcom}
	\end{align} 

\par
If one views disease as the disruption of a process leading to cellular or organismal dysfunction, one natural hypothesis is that SNPs with the greatest potential to disrupt cellular processes might be enriched for disease association. To test this we used both the Wilcoxon rank-sum and Kolmogorov-Smirnov (KS) tests to assay whether the $259$ NHGRI GWAS-annotated SNPs in the GCC were more likely to have high $Q_{ih}$ scores. For both tests, the distribution of $Q_{ih}$ scores for GWAS-associated SNPs were compared to the distribution of non-GWAS SNP scores. 
\par
\begin{figure}[h!]
\centering
\includegraphics[width=0.75\textwidth]{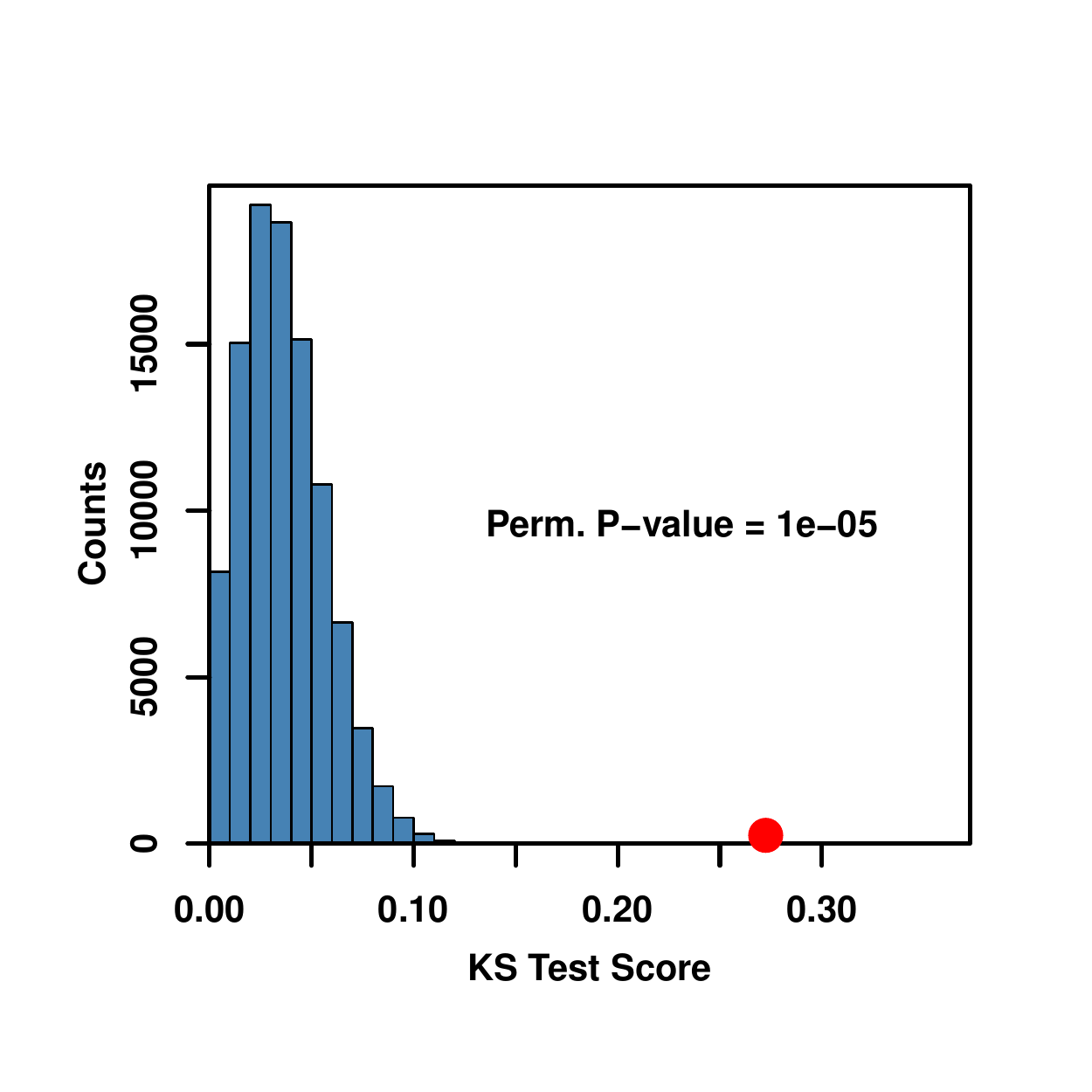}
\caption{Histogram of Kolmogorov-Smirnov test statistics comparing the distribution of $Q_{ih}$ scores for sets of randomly relabeled NHGRI-GWAS/non-GWAS SNPs.The KS test statistic for the true labeling is in red. The permutation p-value associated with the KS test is $P < 10^{-5}$ given $10^5$ permutations. \label{fig:ks_hist}}
\end{figure}
\begin{figure}[h!]
\centering
\includegraphics[width=0.75\textwidth]{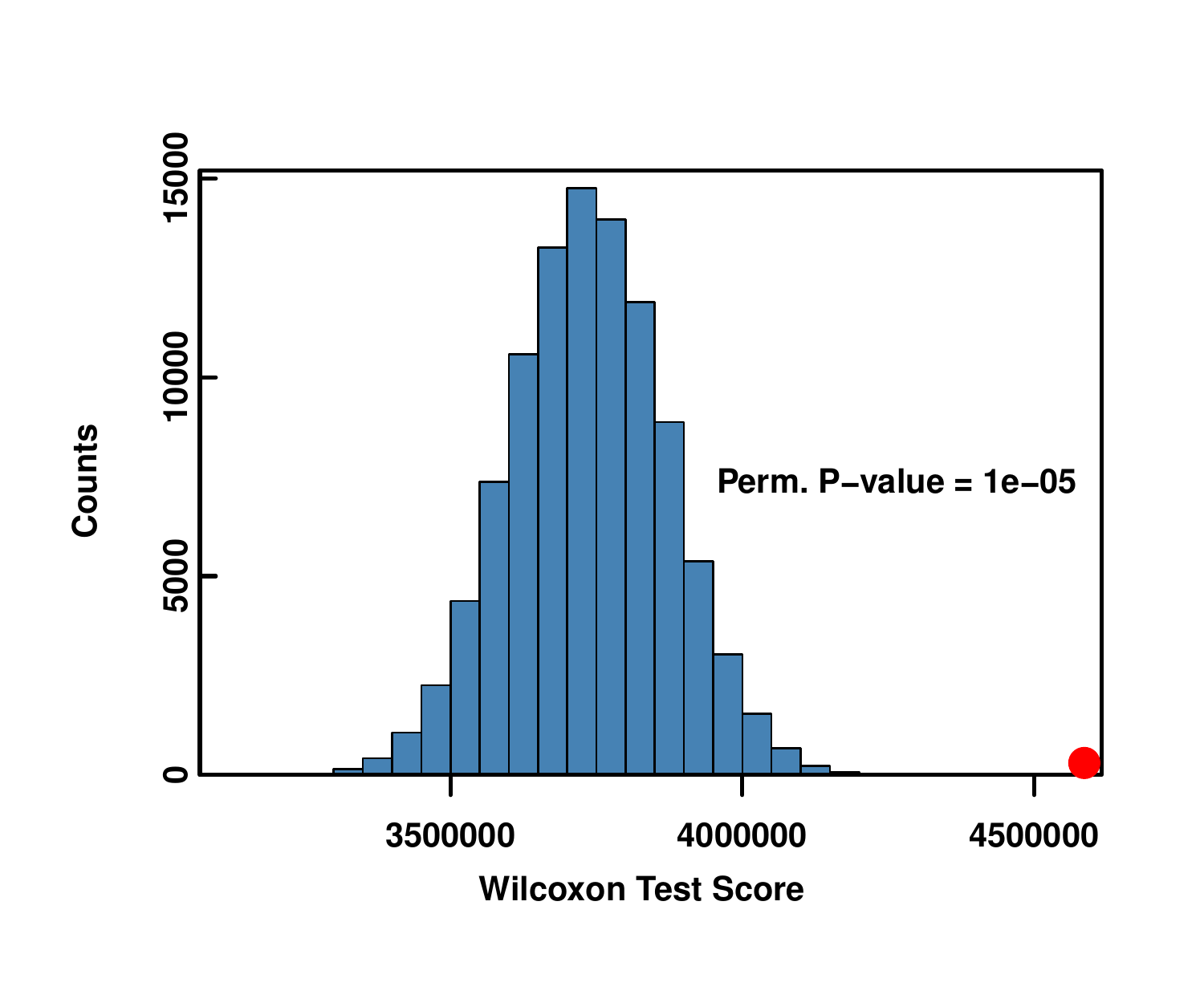}
\caption{\label{fig:wrs_hist} Histogram of Wilcoxon test statistics comparing the distribution of $Q_{ih}$ scores for sets of randomly relabeled NHGRI-GWAS/non-GWAS SNPs. The Wilcoxon test statistic for the true labeling is in red. The permutation p-value associated with the Wilcoxon test is $P < 10^{-5}$ given $10^5$ permutations.}
\end{figure}

To obtain an empirical p-value for these tests, we permuted the GWAS/non-GWAS labels and recalculated the KS and Wilcoxon tests $10^{5}$ times. Histograms of the randomized labels are shown in Figure \ref{fig:ks_hist} and (Supplementary Figure \ref{fig:wrs_hist}). The red dot in the histogram represents the test score with the true labeling. Both tests had highly significant permutation p-values, with $P < 10^{-5}$ for the KS and Wilcoxon tests, indicating that GWAS SNPs were over-represented among SNPs with high core scores. Furthermore, the median core score for the GWAS SNPs was $1.69$ times higher than the median core score for the non-GWAS SNPs. Thus, while global hubs are depleted for GWAS associations with disease, local hubs are significantly enriched for disease associations.

\section*{Discussion}
Genome-wide association studies have searched for genomic variants that influence complex traits, including the development and progression of disease. However, the number of highly-penetrant Mendelian variants that have been found is surprisingly small, with most disease-associated SNPs having a weak phenotypic effect. GWAS studies have also identified many SNPs that do not alter protein coding and have found significant loci that are shared in common across multiple diseases. This body of evidence suggests that in most instances it is not a single genetic variant that leads to disease, but many variants of smaller effect that together can disrupt cellular processes that lead to disease phenotypes. The challenge has been to find these variants of small effect and to place them into a coherent biological context.

\par

We chose to address this problem by analyzing the link between genetic variants and the most immediate phenotypic measure, gene expression. In doing so, we chose not to focus solely on \emph{cis-}acting SNPs, but also to consider \emph{trans-}acting variants. Our motivation was, in part, to try to understand SNPs found through GWAS studies to be associated with phenotypes, but that could not be immediately placed into a functional context. After performing a genome-wide \emph{cis-} and \emph{trans-} eQTL analysis, we identified a large number of many-to-many associations: single SNPs associated with many genes as well as single genes that were significantly associated with many SNPs. To represent those associations, we constructed a bipartite network, one that contains two types of nodes---SNPs and genes---with edges connecting SNPs to the genes with which they were significantly associated. Our analysis of that network led to a number of observations that independently speak to our intuition about disease and the genetic factors that control it.
\par
First is the observation that the highly connected SNPs, the global hubs in the network, are devoid of variants that have been identified as being disease-associated in the hundreds of studies collected in the NHGRI GWAS catalog. While initially surprising, further consideration suggests that this may be the result of negative selection. Since a true hub SNP influences genes across the genome that are involved in many biological processes, highly deleterious variants are likely to significantly disrupt cellular function. In fact, this is the expected impact of a hub---its disruption should lead to the catastrophic collapse of the network.  And so, deleterious SNPs that are network hubs are likely to be lethal or highly debilitating and therefore strongly selected against and quickly swept from the genome. 
\par
Second, we found that SNPs and their target genes form highly connected communities that are enriched for specific biological functions. This too speaks to our inituition and to the evidence about polygenic traits that has accumulated over time. They are not the result of a single SNP that regulates a single gene, but a family of SNPs that together help mediate a group of functionally-related genes.
\par
Third, the enrichment for GWAS disease associations among the high $Q_{ih}$ SNPs has a very simple and intuitive interpretation. The SNPs that are most significantly connected within a particular functionally-related group are those most likely to disrupt that process and therefore be discovered in GWAS analysis. After all, diseases do not develop because the cell's entire functionality collapses, but because specific processes within the cell are disrupted.
\par
What our analysis provides is a new way of exploring \emph{cis-} and \emph{trans-} eQTL analysis and GWAS. What one must do is to consider not only the local effects of genetic variants, but also the complex network of genetic interactions that help regulate phenotypes, including gene expression. 
\par
It also suggests a new way of filtering genes for inclusion in GWAS analysis. Since many disease-associated SNPs appear to be either \emph{cis-}acting or those which are central to functionally-defined communities, one should focus on those SNPs most likely perturb specific biological processes rather than considering the entirety of SNPs in the genome.
\par
\begin{figure}[h!]
\centering
\includegraphics[trim = 1.5cm 1cm 1cm 1cm, clip,width=1\textwidth]{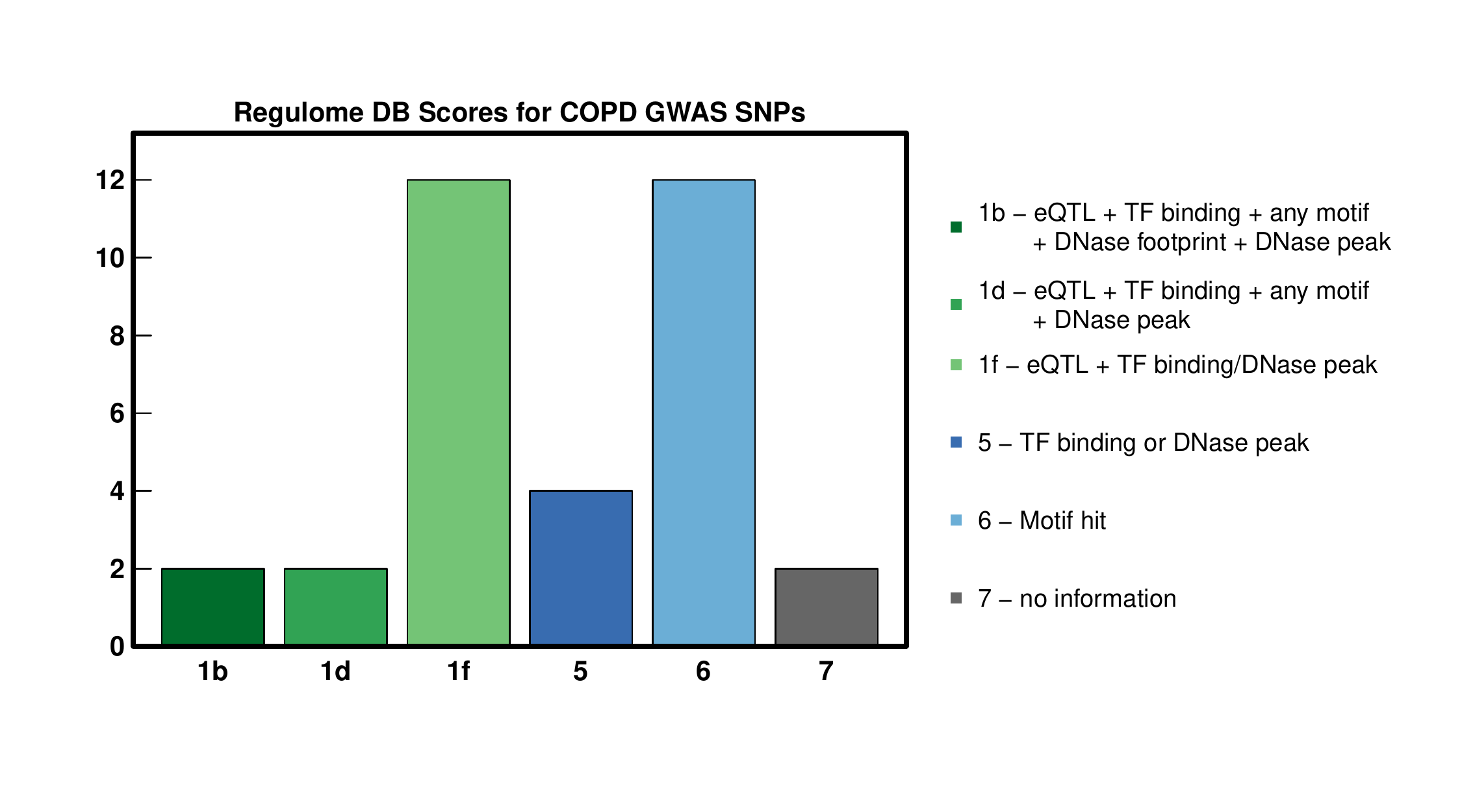}
\caption{Of the 34 SNPs that are eQTLs in the GCC of the LGRC network and also associated with COPD (FDR $< 0.05$), 16 are likely to affect transcription factor (TF) binding and linked to the expression of a target gene (a score of 1b, d, or f), 4 have evidence of TF binding or a DNase peak (a score of 5), and 12 are located in a motif hit (a score of 6) according to RegulomeDB \cite{regulomedb}.     \label{fig:regulomedb_hist}}
\end{figure}

As a way of further assessing the link between GWAS significance and functional perturbation in COPD, we calculated a GWAS-FDR for all SNPs in the GCC of our network that had a reported p-value from a recent GWAS and meta-analysis of COPD \cite{cho2014} (See Supplemental Methods). There were 34 SNPs with an FDR $< 0.05$, and 32 of the 34 had evidence of functional impact according to RegulomeDB \cite{regulomedb}, with 16 SNPs identified as likely to affect transcription factor binding and linked to expression (See Fig.\ \ref{fig:regulomedb_hist}). These 34 SNPs mapped to 4 different communities including Community 30, which contains other COPD-associated SNPs and genes, and is enriched for GO terms describing T cell function and immune response. One of the SNPs in this community likely to affect binding is rs9268528, which is linked by our network to HLA-DRA, HLA-DRB4, and HLA-DRB5; the \emph{cis-}eQTL associations between rs9268528 and both HLA-DRA and HLA-DRB5 have been previously observed in lymphoblastoid cells \cite{montgomery2010transcriptome}. All three HLA genes lie in Community 30 and contribute to the community's enrichment for T cell receptor signaling pathway (GO:0050852) \cite{amigo}.
\par
\begin{figure}[h!]
\centering
\includegraphics[width=0.75\textwidth]{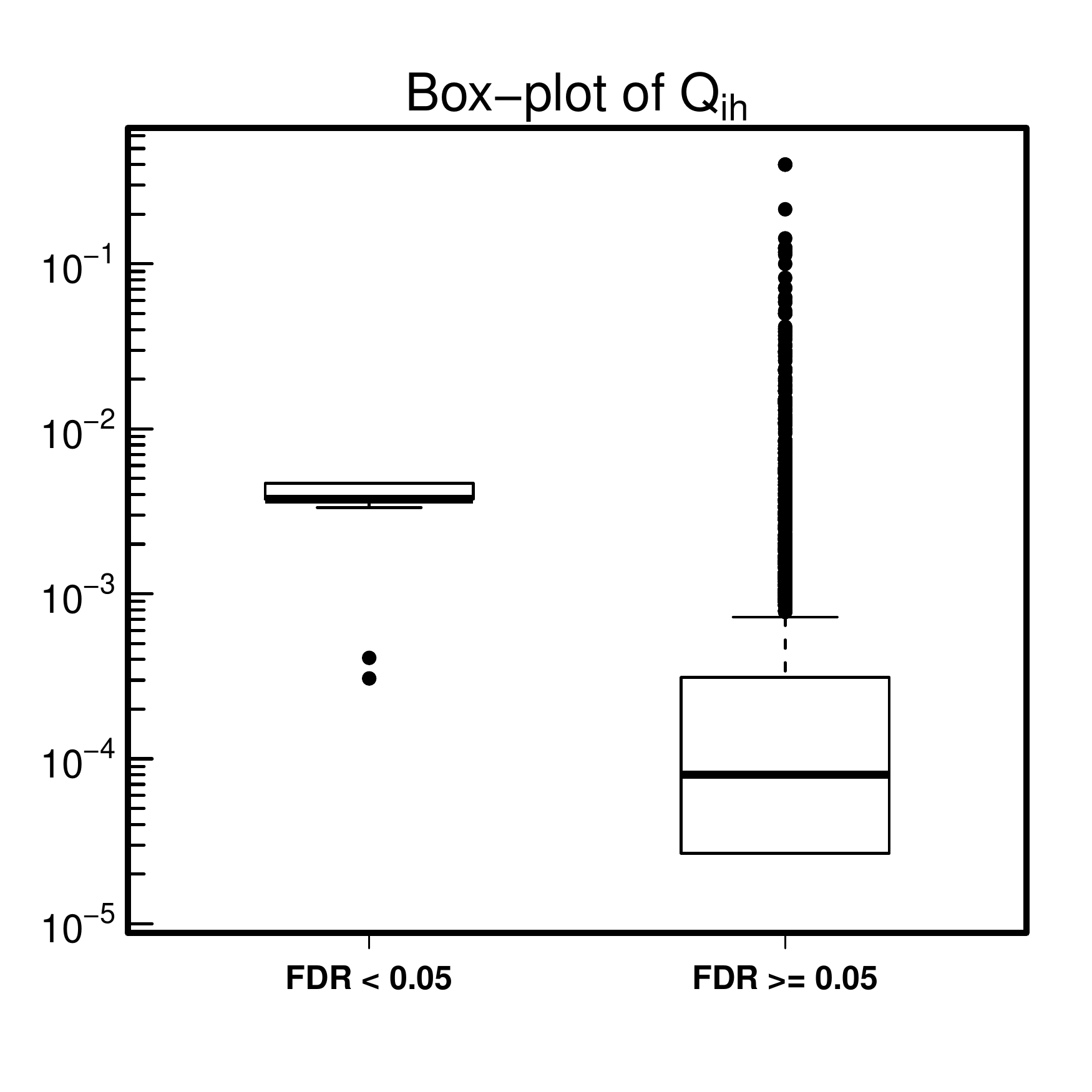}
\caption{The median core scores for the 34 FDR-significant COPD GWAS SNPs (FDR $< 0.05$, left) is 46.7 times higher than the median core score for the non-significant SNPs (FDR $\ge 0.05$, right). \label{fig:gwasbox}}
\end{figure}

To determine the network influence of these 34 SNPs, we compared their core score, $Q_{ih}$ (see Section \ref{sec:community} and Eq. (\ref{eq:coreq})) to the core scores of SNPs with a GWAS-FDR $\ge 0.05$ (See Fig. \ref{fig:gwasbox}). The median $Q_{ih}$ value for the 34 GWAS-FDR significant SNPs was 46.7 times higher than the median for SNPs with an FDR $\ge 0.05$. 
\par
One might note that this analysis was carried out using data on genetic variation and gene expression from the LGRC representing COPD and control lung tissue and question both the generalizability of the results and the use of GWAS-associated disease SNPs from many diseases in the analysis. While these are potentially legitimate concerns, many of the community-based processes we find are not specific to COPD or to the lung but instead are active in nearly all human cell types. 
\par
Although one might expect some processes to change in different disease states, the impact of common variants and the structure of the network is likely to be highly similar. Consequently, although there may be some SNPs whose impact is disease and tissue specific, many are likely to be independent of disease state. This suggests that it may be useful to develop eQTL networks across disease states and tissue types and to explore changes in the overall network and community structure across and between phenotypes due to rare variants and tissue-specific expression.
\par
Validating individual associations in the eQTL network is a difficult challenge. Most eQTL studies limit their validation efforts to downstream effects of high-confidence \emph{cis-}acting eQTLs. The bipartite network presented here captures not only these strong \emph{cis-}eQTLs but also the weak effects of many more \emph{cis-} and \emph{trans-}acting SNPs. So the likelihood that any individual association can be easily validated may not be that great, as it is likely to be of small phenotypic effect and important in only a subset of individuals. However, this is not the point. What is important for the phenotype is not any single SNP-gene association, but the ``mesoscale" organization of genes and SNPs represented by the communities in the network. We believe this intermediate structure better reflects the aggregation of weak genetic effects that contribute to late-onset complex diseases. What we hope to have demonstrated in this manuscript is that the higher order structure, which was not an input to the model, provides insight into a number of aspects of the genetics of polygenic traits consistent with our understanding of how these traits manifest themselves.

\section*{Acknowledgements}
This work was supported by grants 1P01HL105339 (PC, DD, JP, JQ) and 1R01HL111759 (DD, JP, JQ) from the US National Heart Lung Blood Institute of the National Institutes of Health and grant 1R01AI099204 from the US National Institute of Allergy and Infections Disease of the National Institutes of Health (JP, JQ). PC was also supported by grants from the US National Heart, Lung, and Blood Institute (K08 HL102265 and R01 HL124233). Primary data used in this analysis was generated by the Lung Genomics Research Consortium (RC2 HL101715). Previously published COPD GWAS results were supported by COPDGene(R01 HL089856 and R01 HL089897). We thank Drs. Michael Cho and Ed Silverman for their insight and assistance with the COPD GWAS data.

\bibliography{bipartite_copd_refs.bib}


\beginsupplement
\section{Supplemental Methods}
We began by downloading gene expression data from the LGRC web portal (\url{https://www.lung-genomics.org/download/}) representing data from COPD-case and control samples generated by the Lung Genomics Research Consortium (LGRC). This included GCRMA-normalized gene expression data obtained using Agilent-014850 Whole Human Genome 4x44K and Agilent-028004 SurePrint G3 Human GE 8x60K Microarrays. We then obtained matching genotyping data (dbGAP accession phs000624.v1.p1) collected using the Illumina Infinium HD Assays with Human Omni 1Quad and Human Omni 2.5 Quad arrays. All subjects were reported to be of Caucasian descent and were selected based on a variety of parameters including clinical measures associated with diagnosis. Samples that did not meet standards for lack of relatedness as measured using Identity by Descent (IBD) and inbreeding coefficient, $F$, were excluded. Those samples with discordance between reported and genetic sex were not included. Samples missing more than 10\% percent of genotyped SNPs were also removed. SNPs with minor allele frequency (MAF) $< 0.05$ or Hardy Weinberg Equilibrium $< 0.001$ were removed. After all quality controls, 163 samples remained. The COPD GWAS data from a meta-analysis of COPDGene non-Hispanic whites and African-Americans, ECLIPSE, GenKOLS, and NETT/NAS studies was obtained from the authors of \cite{cho2014}.
\par
\subsection{Power-law Fitting}
For each empirical degree distribution, we fit the two parameters for a power-law: the minimum degree at which the power-law behavior starts, $d_{min}$, and the exponent, $\alpha$. A Kolmogorov-Smirnov test was then used to estimate the goodness of fit between 5,000 randomly generated power-law distributed synthetic data sets given $d_{min}$ and $\alpha$ and their corresponding power-law fit. The probability, $P_{pl}$, that a degree distribution follows a power-law with $d_{min}$ and $\alpha$ is then the fraction of times a synthetic data set has a KS statistic larger than that of the true test. For both the SNP and gene degree distributions, $P_{pl}$ was calculated using the 5,000 goodness of fit values (code for the parameter estimation, goodness of fit and probability estimation was obtained from \url{http://tuvalu.santafe.edu/~aaronc/powerlaws/}). 

\begin{figure}[h!]
\centering
\includegraphics[width=0.75\textwidth]{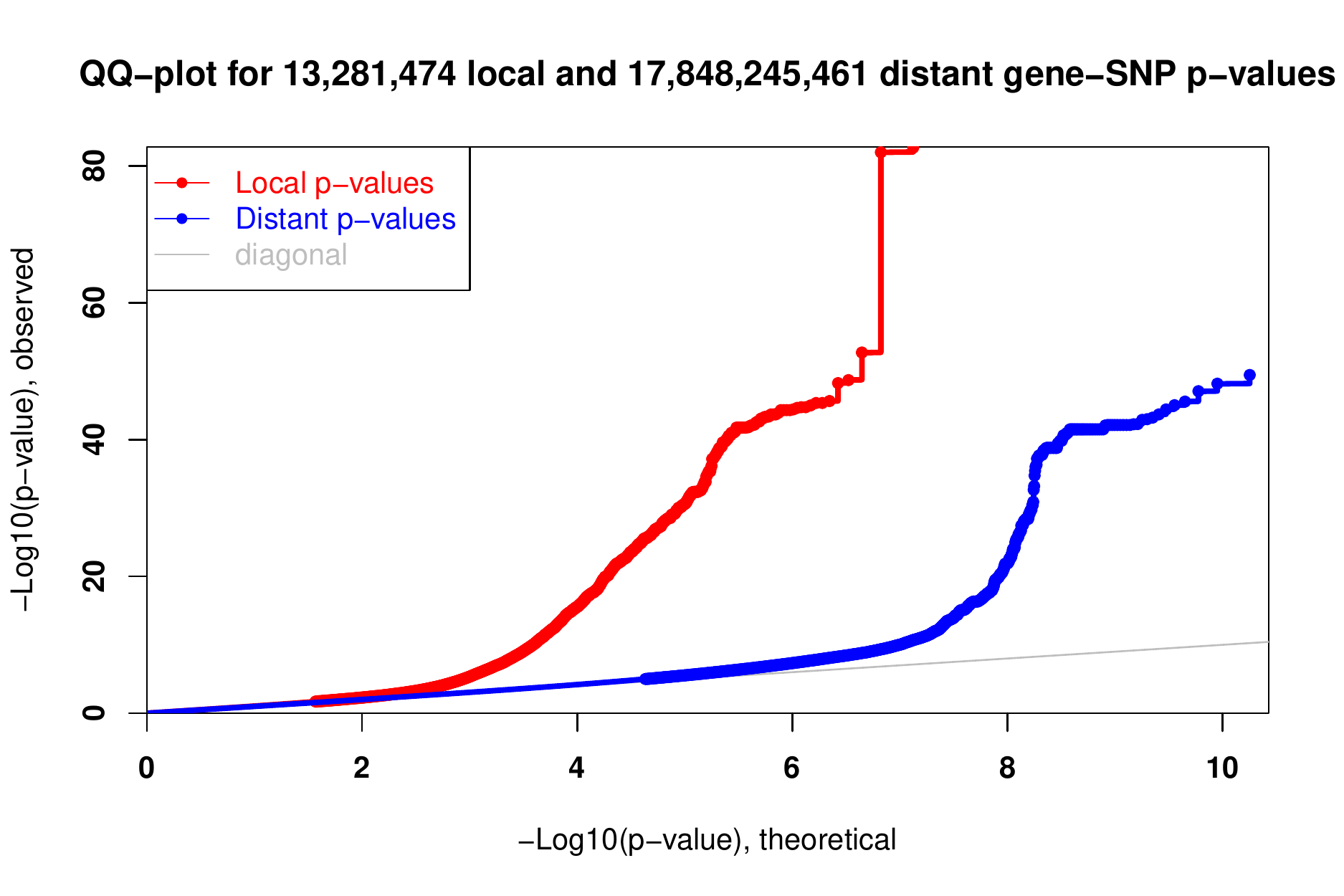}
\caption{Quantile-quantile plot for \emph{cis-} and \emph{trans-}eQTL associations. \label{fig:qqplot}}
\end{figure}
%
\end{document}